\documentclass{aastex63}
\submitjournal{ApJL}
\usepackage{amsmath}
\newcommand{\MyFigA}{\textcolor[rgb]{0.00,0.00,1.00}{1}}
\newcommand{\MyFigB}{\textcolor[rgb]{0.00,0.00,1.00}{2}}
\newcommand{\MyFigC}{\textcolor[rgb]{0.00,0.00,1.00}{3}}

\newcommand{\R}{\textcolor[rgb]{0.20,0.80,0.00}}
\shorttitle{Sample article}
\shortauthors{Schwarz et al.}
\begin{document}
\title{Gamma-ray Burst Spectrum with a Time-dependent Injection Rate of High-energy Electrons}
\correspondingauthor{Da-Bin Lin}
\email{lindabin@gxu.edu.cn}
\author{Kuan Liu}
\author{Da-Bin Lin}
\author{Kai Wang}
\author{Li Zhou}
\author{Xiang-Gao Wang}
\author{En-Wei Liang}
\affiliation{Guangxi Key Laboratory for Relativistic Astrophysics, School of Physical Science and Technology, Guangxi University, Nanning 530004, China}
\begin{abstract}
Although the physical origin of prompt emission in gamma-ray bursts (GRBs) remains inconclusive,
previous studies have considered the synchrotron radiation of
relativistic electrons as a promising mechanism.
These works usually adopted a invariable injection rate of electrons ($Q$)
which may be discordant with that in a Poynting-flux dominated jet.
In a Poynting-flux dominated jet (e.g., ICMART model \citealp{Zhang_B-2011-Yan_H-ApJ.726.90Z}),
the number of magnetic reconnections occurred simultaneously
may grow rapidly with time and results in an increase of $Q$ with time.
This paper is dedicated to study the synchrotron radiation spectrum in this scenario.
It is found that the radiation spectrum would obviously get harder
if an increasing $Q$ is adopted
and a Band-like radiation spectrum can be obtained
if the increase of $Q$ is fast enough.
The latter is related to the fact that
a bump-shape rather than a power-law spectrum appears
in the low-energy regime of the obtained electron spectrum.
This effect can strongly harden the low-energy radiation spectrum.
It indicates that an increasing $Q$ can help to alleviate
the ``fast-cooling problem'' of synchrotron radiation for GRBs.
Our studies also reveal that
a Poynting-flux dominated jet with a large emission radius,
a small length of the magnetic reconnection region,
or a low-minimum energy of injected electron would
prefer to form a Band-like radiation spectrum.
We suggest that the Band spectrum found in GRBs may be
the synchrotron emission of the electrons
with a bump-shape distribution in its low-energy regime.
\end{abstract}

\section{Introduction}
Although the rich observational data for the prompt emission phase of gamma-ray bursts (GRBs) has been detected, its radiation mechanism remains unclear to astrophysicists.
For the majority of GRBs, the energy spectra of prompt emission can be well fitted with
a smoothly connected broken power-law, which is known as Band function (\citealp{Band_D-1993-Matteson_J-ApJ.413.281B}).
The typical parameters of Band function are $\alpha\sim-0.8$, $\beta\sim-2.3$, and $E_{\rm p}\sim 250$~keV
with $\alpha$, $\beta$ and $E_{\rm p}$ being the low-energy photon spectral index, high-energy photon spectral index, and peak photon energy of $\nu F_{\nu}$ spectrum (\citealp{Preece_RD-2000-Briggs_MS-ApJS-126-19P}).
Owing to the non-thermal feature of the radiation spectrum, it is commonly believed that
the synchrotron radiation may be the main radiation mechanism being responsible to the prompt emission (\citealp{Katz_JI-1994-ApJ-432L-107K}; \citealp{Meszaros_p-1994-Rees_MJ-ApJ-432-181M}; \citealp{Daigne_F-1998-Mochkovitch_R-MNRAS-296-275D}; \citealp{Guiriec_S-2015-Kouveliotou_C-ApJ.807.148G}; \citealp{Burgess_JM-2017-ifs..confE..74B}; \citealp{Oganesyan_G-2017-Nava_L-ApJ...846..137O}; \citealp{Ravasio_ME-2018-Oganesyan_G-A&A...613A..16R}).
However, there is a famous ``fast-cooling problem'' of the synchrotron radiation for the prompt emission (\citealp{Preece_RD-1998-Briggs_MS-ApJ-506L-23P}; \citealp{Ghisellini_G-2000-Celotti_A-MNRA-313L-1G}; \citealp{Kumar_P-2008-McMahon_E-MNRAS.384.33K}).
To produce a synchrotron radiation spectrum with an $E_{\rm p}$ around sub-MeV,
the magnetic field in the emission region should be strong enough,
which may put the electrons in the fast-cooling regime.
In this regime, it is believed that the photon spectral index should be
$-3/2$,
which is too soft compared with the typical Band function (\citealp{Sari_R-1998-Piran_T-ApJ.497L.17S}; \citealp{Nava_L-2011-Ghirlanda_G-A&A-530A-21N}; \citealp{Zhang_BB-2011-Zhang_B-ApJ.730.141Z}).
There are numerous studies proposed to harden the low-energy radiation spectrum of the synchrotron radiation, such as, considering a decaying magnetic field in the emission region (\citealp{Peer_A-2006-Zhang_B-ApJ.653.454P}; \citealp{Uhm_ZL-2014-Zhang_B-NatPh.10.351U}; \citealp{Zhao_XH-2014-Li_Z-ApJ...780...12Z}), introducing a slow heating in the downstream turbulence of the internal shock (\citealp{Asano_K-2009-Terasawa_T-ApJ...705.1714A}), involving the inverse Compton scattering cooling effect in the Klein-Nishina regime (\citealp{Derishev_EV-2001-Kocharovsky_VV-A&A.372.1071D}; \citealp{Nakar_E-2009-Ando_S-ApJ...703..675N}), or considering a marginally fast cooling regime (\citealp{Daigne_F-2011-Bosnjak_Z-A&A.526A.110D}; \citealp{Beniamini_P-2018-Barniol_DR-MNRAS.476.1785B}).

On the other hand, the absence of thermal component of some spectra indicates that a fraction of GRBs is somewhat Poynting-flux dominated (\citealp{Zhang_B-2009-Peer_A-ApJ.700L.65Z}; \citealp{Guiriec_S-2011-Connaughton_V-ApJ.727L.33G}; \citealp{Axelsson_M-2012-Baldini_L-ApJ.757L.31A}).
In a Poynting-flux dominated jet, the magnetic reconnection plays an important role in energy dissipation (\citealp{Zhang_B-2005-Kobayashi_S-ApJ.628.315Z}).
It has been shown that the native Sweet-Parker reconnection process proceeds very slowly in the astrophysical environment (\citealp{Sweet_PA-1958-IAUS-6-123S}; \citealp{Parker_EN-1957-JG-62-509P}).
\cite{Lazarian_A-1999-Vishniac_ET-ApJ-517-700L} indicated that the magnetic turbulence,
which is naturally present in most of astrophysical systems,
could enhance the speed of the reconnection and trigger lots of independent reconnections to occur simultaneously.
This behavior has been verified in the three-dimensional numerical simulations of \cite{Kowal_G-2009-Lazarian_A-ApJ-700-63K}.
According to this kind of fast reconnection process, \cite{Zhang_B-2011-Yan_H-ApJ.726.90Z} proposed the internal-collision-induced magnetic reconnection and turbulence (ICMART) model to explain
the origin of the prompt emission of GRBs.
They envisaged an avalanche-like magnetic dissipation process
and pointed out that the reconnection events would be exponentially increasing
until most of magnetic energy is released (e.g., \citealp{Zhang_B-2014-Zhang_B-ApJ.782.92Z}).
That is to say, in this scenario,
the injection rate of electrons would increase rapidly with time.
This is apparently different from what is usually adopted in the standard fast-cooling pattern.
An increasing injection rate can raise the number of high-energy electrons
in the electron spectrum and thus may harden the corresponding radiation spectrum.
This may help to alleviate the fast-cooling problem to some degree.
In this paper, we study the effect of an increasing injection rate of electrons
on the synchrotron radiation spectrum.

The structure of this paper is organized as follows.
In Section~\ref{Sec:model}, we delineate our prescription for the injection rate of electrons
and general picture of our calculations.
In Section~\ref{Sec:result},
we unfold the effect of an increasing injection rate of electrons on the radiation spectrum.
The conclusions and discussions are presented in Section~\ref{Sec:Conclusion}.

\section{Model}\label{Sec:model}
\subsection{Prescription for the Injection Rate of Electrons}
In the synchrotron radiation mechanism, the GRB prompt emission comes from a group of electrons.
In this paper, the instantaneous electron spectrum at time $t'$ is denoted as $n'_{\rm e}(\gamma'_{\rm e}, t')$,
where $n'_{\rm e}(\gamma'_{\rm e}, t')d\gamma'_{\rm e}$ is the number of electrons in $[\gamma'_{\rm e}, \gamma'_{\rm e}+d\gamma'_{\rm e}]$ with $\gamma'_{\rm e}$ being the Lorentz factor of electrons.
Throughout this paper, the primed quantities are measured in the comoving frame of the jet shell.
The evolution of $n'_{\rm e}(\gamma'_{\rm e}, t')$ can be described as
\begin{equation}\label{EQ:Continuity}
\frac{\partial n'_{\rm e}}{\partial t^\prime}+
\frac{\partial ({\dot{\gamma}}'_{\rm e}n'_{\rm e})}{\partial\gamma^\prime_{\rm e}}=Q'\left(\gamma_{\rm e}^\prime,t^\prime\right),
\end{equation}
where $\dot{\gamma}'_{\rm e}=d {\gamma}'_e/dt'$ is the cooling rate of electrons and
$Q'(\gamma_{\rm e}^\prime,t^\prime)$ is the injection rate of electrons at ${\gamma}'_{\rm e}$ and $t'$.
In previous studies, $Q'\left(\gamma_{\rm e}^\prime,t^\prime\right)=A'{\gamma'}_e^{-p}$ with a constant $A'$
is usually adopted to describe the injection rate of electrons.
In this paper, however, we point out that $A'$ may be a function of $t'$,
especially for that in a Poynting-flux dominated jet.

In a Poynting-flux dominated jet,
the outflow may be composed of wound-up magnetic field lines
(\citealp{Coroniti_EV-1990-ApJ...349..538C}; \citealp{Spruit_HC-2001-Daigne_F-A&A...369..694S}).
The magnetic field lines in the jet may be distorted due to some instabilities
or internal collisions (e.g., ICMART, \citealp{Zhang_B-2011-Yan_H-ApJ.726.90Z}).
At a certain point, it would reach a critical condition to
allow fast reconnections and turbulence to occur.
The turbulence may further distort field lines and induce additional reconnections,
resulting in a catastrophically discharge of the magnetic energy in the jet.
In this situation, the number of reconnections (same as $Q$) would increase rapidly with time.
For simplicity, we assume that the reconnection length $l'_{\rm rec}$ and velocity $v'_{\rm rec}$ are the same for each reconnection.
In highly magnetized fluid, $v'_{\rm rec}$ may reach a value of $0.1c$ with $c$ being the light velocity (\citealp{Lyubarsky_YE-2005-MNRAS.358.113L}).
Then, the duration of each reconnection $t'_{\rm rec}$ is defined as $t'_{\rm rec}\backsimeq l'_{\rm rec}/v'_{\rm rec}$ (here the thickness of the bunch of magnetic fields has been approximated to $l'_{\rm rec}$).
The total number of reconnections ${N'_{{\rm{rec}},{\rm{tot}}}}$ is simply written as the ratio between the total dissipated volume and the volume of the region affected by each reconnection, which is estimated as $l '^3_{\rm rec}$.
Within ${1}/{\Gamma}$ cone, one has ${N'_{{\rm{rec}},{\rm{tot}}}} = R_0^2\Delta '\delta \Omega /{l'}_{{\rm{rec}}}^3$ with $\Delta '$ being the thickness of the jet shell and $\delta \Omega=2\pi \left[ {1 - \cos \left( {1/\Gamma } \right)} \right]$.
With $\Delta'= R_0/\Gamma $, one has
\begin{equation}\label{Eq:RecTotNum}
N'_{\rm rec,tot}\simeq\frac{{\pi R_0^3}}{{{\Gamma ^3}{l'}_{{\rm{rec}}}^3}},
\end{equation}
where ${R_0}$ is the radius of the emission region from the central engine and $\Gamma=300$ is the bulk Lorentz factor.

The exact growth behavior of reconnection number is not fully clear yet.
\cite{Zhang_B-2014-Zhang_B-ApJ.782.92Z} invoked an exponential growth pattern in their work.
They imagine that one magnetic reconnection would trigger two reconnections and these two reconnections
would trigger four reconnections in next generation.
In this scenario, the number of reconnections at $t'$ could be described as $N_{\rm rec}(t')\propto 2^{t'/t'_{\rm rec}}$.
Since the total number of reconnections is fixed, the real growth rate of reconnection number may be slower than that of the exponential case in the end.
Due to the lack of fully numerical simulation, we adopt two simple growth models in this paper, i.e., exponential growth and power-law growth,
\begin{equation}\label{Eq:RecGrow}
N'_{\rm rec}(t')=\left\{{\begin{array}{l}{{\rm exp}({t'/t'_{\rm rec}}),}\\ {(1+t'/t'_{\rm rec})^\xi,}\end{array}}\right.
\end{equation}
where $N'_{\rm rec}(t')dt'$ describes the number of reconnections occured in the period of $[t', t'+dt']$.
The end time $t'_{\rm end}$ of an avalanche-like event can be estimated with ${N'_{{\rm{rec}},{\rm{tot}}}} = \int_0^{t{'_{{\rm{end}}}}} {{N'_{{\rm{rec}}}}(t')dt'}$, where $t'=0$ is the time when jet reaches at $R_0$.
Considering the energy conservation, the strength of magnetic field is related to the energy
dissipated by reconnections, i.e., $B'(t')=\sqrt{L_B(t')/(\Gamma^2R_0^2c)}$,
where $L_B(t')=L_{B}(t'=0)-E_{\rm dis}\int_0^{t'} {{N'_{{\rm{rec}}}}(s')ds'}/{N'_{{\rm{rec}},{\rm{tot}}}}$
with $L_{B}(t'=0)=L_{\rm jet}\sigma_0/(\sigma_0+1)$ is the initial magnetic energy at $t'$,
$E_{\rm dis}=L_{\rm jet}(\sigma_0-\sigma_{\rm end})/(\sigma_0+1)$ is the total dissipated magnetic energy,
$L_{\rm jet}$ is the jet power,
and $\sigma_0=100$ and $\sigma_{\rm end}=1$ are the initial and final magnetization parameters, respectively.

Each of these reconnections is modeled as a same plasma accelerator which follows the first Fermi acceleration mechanism.
We assume that the accelerated electrons distribute uniformly in the jet shell.
The energy spectrum of injected electrons is described with a power-law function, i.e.,
\begin{equation}\label{Eq:Q}
Q'(\gamma'_{\rm e},t')=A'(t')\left\{ {\begin{array}{*{20}{c}}{({\gamma}'_{\rm e}/{\gamma}'_{\rm m})^{-p},}&{{\gamma}'_{\rm m}\leqslant\gamma'_{\rm e}\leqslant\gamma'_{\rm max},}\\ {0,}&{\rm others,}\end{array}} \right.
\end{equation}
where $p\simeq2.8$ is a power-law index, $A'(t')=A'_0N_{\rm rec}(t')$ with ${A'_0}{\gamma'}_{\rm m}^2/(2 - p)$ being the energy used to accelerate electrons in each reconnection,
and $\gamma'_{\rm m}$ and $\gamma'_{\rm max}\simeq10^8\left(B'/1{\rm Gs}\right)$ (\citealp{Dai_ZG-1999-Lu_T-ApJ.519L.155D}; \citealp{Huang_YF-2000-Gou_LJ-ApJ.543.90H}) are
the minimum and maximum Lorentz factor of injected electrons, respectively.
We assume that a half of the energy dissipated in one magnetic reconnection is used to accelerate electrons.
Hereafter, the situations with an exponential and a power-law increasing injection rate are represented with EXP-$Q$ and PL-$Q$, respectively.
While the accelerated electrons are moving in the magnetic field $B^\prime$, they would lose their energy via synchrotron radiation and adiabatic expansion cooling (\citealp{Rybicki_GB-1979-Lightman_AP-rpa..book.....R}; \citealp{Uhm_ZL-2012-Zhang_B-ApJ.761.147U}), i.e.,
\begin{equation}\label{EQ:SynCool}
\ {\dot{\gamma}}'_{\rm e, syn}=-\frac{\sigma_T{\gamma'_{\rm e}}^2{B'}^2}{6\pi m_ec}
\end{equation}
and
\begin{equation}\label{EQ:ExpCool}
 {\dot{\gamma}}'_{\rm e, adi}=\frac{1}{3}\gamma'_{\rm e}\frac{d{\rm ln}n'}{dt'}=-\frac{2}{3}\frac{\gamma'_{\rm e}}{R}\frac{dR}{dt'},
\end{equation}
where $\sigma_{\rm T}$ and $m_{\rm e}$ are the Thomson cross-section and electron mass, respectively.
Then $\dot{\gamma}'_{\rm e}=\ {\dot{\gamma}}'_{\rm e, syn}+ {\dot{\gamma}}'_{\rm e, adi}$.
$n'$ is the number density of electrons in the shell, which is approximated as $n'\propto R^{-2}$ for an expanding shell.

The major task of the presented work is to solve Equation~(\ref{EQ:Continuity}) for $n'_{\rm e}$ at different time $t'$.
In our calculations, the 4th-order Runge-Kutta method is used.
In addition, an appropriate time step $\Delta t' < \min \left\{ {\left| {\Delta {{\gamma '}_{\rm{e}}}/{{\dot \gamma '}_{{\rm{e}},{\rm{tot}}}}} \right|} \right\}$ is adopted in our calculations,
where $\Delta \gamma'_{\rm e}$ is the width of our adopted energy grids for electrons
(e.g., see the appendix~A of \citealp{Geng_JJ-2018-Huang_YF-ApJS..234....3G}).

\subsection{Synchrotron Emission}
The spectral power of synchrotron radiation at a given frequency $\nu'$ and $t'$
in the unit solid angle of the jet shell can be described as
\begin{equation}\label{EQ:SynPower}
P'(\nu',t')=\frac{\sqrt{3}q_{\rm e}^3B'(t')}{m_{\rm e}c^2}\int_{0}^{\gamma'_{\rm max}}F\left(\frac{\nu'}{\nu'_{\rm c}}\right)\frac{n'_{\rm e}(\gamma_e', t')}{\delta \Omega}d\gamma'_{\rm e},
\end{equation}
where $q_{\rm e}$ is the electron charge,
$F(x)=x\int_{x}^{+\infty}K_{5/3}(k)dk$ with $K_{5/3}(k)$ being the modified Bessel function of 5/3 order,
and $\nu'_{\rm c}=3q_{\rm e}B'(t')\gamma^2_e/(4\pi m_{\rm e}c)$.
The observed flux density at the observer time $t_{\rm{obs}}$ and frequency $\nu$ is calculated with
\begin{equation}\label{EQ:ObsFlux}
F_{\nu}(t_{\rm{obs}})={\int}\kern-15pt\int\limits_{\rm (EATS)}{\frac{(1+z)P'\left(\nu\frac{1+z}{D},t'\right)D^3}{4\pi d^2_{\rm L}}d \Omega},
\end{equation}
where EATS is the equal-arrival time surface corresponding to the observer time $t_{\rm obs}$,
$D = 1/\left [ \Gamma \left(1-\beta \cos\theta\right)\right ]$ is the Doppler factor,
$\beta=\sqrt{1-1/\Gamma^2}$ is the dimensionless velocity,
$\nu$ is the observed photon frequency,
and $d_{\rm L}$ is the luminosity distance at the cosmological redshift $z=1$.
Then, the observed flux at $t_{\rm obs}$ is $F({t_{{\rm{obs}}}}) = \int_0^{ + \infty } {{F_\nu }({t_{{\rm{obs}}}})d\nu } $.
The observed time $t_{\rm{obs}}$ of a photon emitted from $R$ and $\theta$ can be estimated as follows:
\begin{equation}\label{EQ:ObsTime}
t_{\rm obs}=\left(\int_{R_0}^{R}\frac{dr}{c\beta}-\frac{R{\rm cos}\theta-R_0}{c}\right)(1+z),
\end{equation}
where the observed time of a photon emitted from $R_0$ and $\theta=0$ is set as $t_{\rm obs}=0$.
An assumption of an on-axis observer is adopted in our work.

\section{Results with an Increasing $Q$}\label{Sec:result}

In this section, we investigate the effects of an increasing $Q$ on the synchrotron radiation spectrum and electron spectrum.
We first study the case with an EXP-$Q$, $R_0=10^{16}$ cm, $l'_{\rm rec}=10^{11}$~cm, $\gamma'_{\rm m}=3\times 10^4$, and $L_{\rm jet}=10^{52}\, \rm erg\cdot s^{-1}$.
In this case, the end time $t'_{\rm end}$ of electron injection is $840$~s,
corresponding to the observer time of $t_{\rm obs}=1.4$~s.
The obtained light curve of flux is shown in the left panel of Figure~{\MyFigA}.
Then, we plot the radiation spectra at $t_{\rm obs}=1$~s (red line), 1.2~s (blue line), 1.4~s (black line) in the middle panel of this figure.
One can find that
the obtained radiation spectra in the low-energy regime are very different from
the standard fast-cooling pattern,
i.e., $F_{\nu}\propto \nu^{-1/2}$,
which is shown with the gray dashed line in this panel.
This result reveals that an increasing $Q$ may help to alleviate
the ``fast-cooling problem'' of the synchrotron radiation.
To better comparison, we also plot the radiation spectrum of $F_{\nu}\propto \nu^{0.2}$
with green dashed line in this panel.
Here, $F_{\nu}\propto \nu^{0.2}$ sketches the low-energy spectral shape of Band function with $\alpha=-0.8$.
One can easily find that our radiation spectrum is very close to the Band function spectrum in its low-energy regime.
It suggests that the situations with an increasing $Q$ may produce a Band-like radiation spectrum in the synchrotron radiation scenario.
For better understanding our radiation spectra,
we plot the electron spectra in the jet flow located at $\theta=0$ and
observed at $t_{\rm obs}=1$~s (red line), $1.2$~s (blue line), $1.4$~s (black line)
in the right panel of Figure~{\MyFigA}.
In this panel, the standard electron spectra in the fast-cooling regime
and the electron spectra required to produce $F_{\nu}\propto \nu^{0.2}$
are also plotted with the gray and green dashed lines, respectively.
One can see that the electron spectrum with an increasing $Q$ is very different from the standard one.
What is more, a bump shape rather than a power-law shape
appears in the low-energy regime of our electron spectra.
These are our main findings in the present work.
The situations with a PL-$Q$ are also studied.
The obtained light curves (left panels), radiation spectra (middle panels), and electron spectra (right panels) are shown in Figure~{\MyFigB},
where the value of $\xi=2$ ($\xi=4$) is adopted in the upper (bottom) panels.
It can be found that the radiation and electron spectra
are also very different from the standard fast cooling patterns,
especially for the case with $\xi=4$.
By comparing the results presented in the upper panels ($\xi=2$) with those in the bottom panels ($\xi=4$), one can find that a faster increase $Q$ would produce a greater deviation of the radiation/electron spectrum
from the standard fast-cooling one.

We next study the effects of the magnetic reconnection length $l'_{\rm rec}$, the initial emission radius $R_0$, and the minimum Lorentz factor $\gamma'_{\rm m}$ of injected electrons on the radiation spectrum.
The low-energy photon spectral index $\alpha$ of the radiation spectrum is our main focus.
Then, the Band function is used to fit the obtained photon spectrum
at the peak time of the light curve.
With our fits, we can obtain the value of $\alpha$.
The dependence of $\alpha$ on $l'_{\rm rec}$ (left panel), $R_0$ (middle panel), and $\gamma'_{\rm m}$ (right panel) is shown in Figure~{\MyFigC},
where the situations by adopting an EXP-$Q$, a PL-$Q$ with $\xi=2$, and a PL-$Q$ with $\xi=4$
are shown with black, red, and blue lines, respectively.
In our calculations, the dependence of $\alpha$ on a parameter (e.g., $l'_{\rm rec}$) is obtained by changing the value of this parameter and
keeping the other two (e.g., $R_0$ and $\gamma'_{\rm m}$) frozen at their fiducial value,
where $l'_{\rm rec}=10^{11}$~cm, $R_0=10^{16}$~cm, $\gamma'_{\rm m}=2\times10^4$, and $L_{\rm jet}=10^{52}\, \rm erg\cdot s^{-1}$ are set as their fiducial values, respectively.
Figure~{\MyFigC} shows that the value of $\alpha$ increases from $\sim -1.5$ to $\sim -0.7$ by decreasing the $l'_{\rm rec}$ or $\gamma'_{\rm m}$, or by increasing the $R_0$.
The relations of $\alpha-l'_{\rm rec}$, $\alpha-R_0$, and $\alpha-\gamma'_m$
can be understood as follows.
By decreasing $l'_{\rm rec}$, the variable timescale of $Q$ decreases.
Then, the dependence of $\alpha$ on $l'_{\rm rec}$ reveals that the increment rate of $Q$ would significantly affect $\alpha$.
The magnetic strength in the emission region
is associated with $R_0$ by $B'=L_{\rm jet}/(\Gamma^2R_0^2c)$.
A high value of $R_0$ would lead to a low magnetic strength $B'$ in the emission region.
A low value of $B'$ or $\gamma'_{\rm m}$ would significantly weaken the synchrotron cooling of electrons and
thus harden the low-energy radiation/electron spectrum.
Since the $B'$ is also related to the jet power, we also study the dependence of $\alpha$ on radius $R_0$ with $L_{\rm jet}=10^{50}\, \rm erg\cdot s^{-1}$,
which is shown in the middle panel of Figure~{\MyFigC} with dashed lines .
One can find that the low-energy radiation spectra are generally hard in this situation.
In summary, a Band-like radiation spectrum is easier to emerge in a jet with low $L_{\rm jet}$, $l'_{\rm rec}$, $\gamma'_{\rm m}$, or with high $R_0$.

The results obtained with an increasing $Q$ can be understood as follows.
For our studied cases, the cooling timescales of most electrons are short compared with the dynamical time $t'$, except those with ${\gamma}'_{\rm e}\sim {\gamma '_{\rm{c}}} = 6\pi {m_e}c/({\sigma _T}{B'^2}t')$.
For ${\gamma}'_{\rm e}>{\gamma '_{\rm{c}}}$, Equation~(\ref{EQ:Continuity}) with a constant $Q$ can be reduced to
\begin{equation}\label{Eq:Fast_ne}
\frac{\partial ({\dot{\gamma}}'_{\rm e}n_{\rm e})}{\partial\gamma^\prime_{\rm e}}=Q\left(\gamma_{\rm e}^\prime,t^\prime\right),
\end{equation}
which leads to the well-known fast-cooling electron spectrum $n'_{\rm e}\propto {\gamma'}_{\rm e}^{-2}$ in ${\gamma '_{\rm{c}}}<{\gamma}'_{\rm e}<{\gamma '_{\rm{m}}}$.
However, the above argument relies on a crucial condition that a constant $Q$ is adopted.
One can imagine that if a time-dependent $Q$ is adopted, the electron spectrum in ${\gamma '_{\rm{c}}}<{\gamma}'_{\rm e}<{\gamma '_{\rm{m}}}$ would be different from the standard one.
If an increase (a decrease) $Q$ is adopted,
the electron spectrum in ${\gamma '_{\rm{c}}}<{\gamma}'_{\rm e}<{\gamma '_{\rm{m}}}$
would be shallower (steeper) than $n'_{\rm e}\propto {\gamma'}_{\rm e}^{-2}$.
If an increasing $Q$ with a short timescale is adopted,
the deviation of our electron spectrum from the standard one would be strong.
Correspondingly, a harder radiation spectrum can be obtained.
What is more, if the increase of $Q$ is fast enough,
there would be a bump-shape rather than a power-law electron spectrum in ${\gamma '_{\rm{c}}}<{\gamma}'_{\rm e}<{\gamma '_{\rm{m}}}$.
This behavior can be found in the right panels of Figures~{\MyFigA} and {\MyFigB}.
With a bump-shape electron spectrum,
the synchrotron radiation spectrum in its low-energy regime
would quickly fall into $F_{\nu}\propto\nu^{1/3}$ as $\nu$ decreases.
Here, $F_{\nu}\propto\nu^{1/3}$ is the synchrotron radiation spectrum of a single electron in its low-energy regime.
Therefore,
a Band-like radiation spectrum with $F_{\nu}\propto\nu^{\sim 0.2}$ in its low-energy regime
rather than the standard fast-cooling pattern
is presented in our model.

\section{Conclusions and Discussion}\label{Sec:Conclusion}
The non-thermal feature of the prompt emission in GRBs
implies that the synchrotron radiation of electrons may be
the main radiation mechanism behind the prompt emission.
However, it may confront the so-called ``fast-cooling problem''
in explaining the prompt emission with the standard synchrotron emission model.
In this paper, we point out that an injection rate $Q$ that increases with time
may help to alleviate the ``fast-cooling problem'' of the synchrotron radiation mechanism.
With an increasing $Q$,
our obtained radiation spectrum get harder in its low-energy regime compared with the standard fast cooling one.
If the increase of $Q$ is fast enough,
a Band-like radiation spectrum can be obtained.
This effect is related to the fact that
with an increasing $Q$,
the obtained electron spectrum becomes shallow in the low-energy regime
compared with the standard fast cooling one.
Furthermore, a bump-shape spectrum will appear in its low-energy regime if the increase of $Q$ is fast enough.
This effect can help to alleviate the ``fast-cooling problem'' of the synchrotron radiation to some degree.
Our studies also reveal that a Poynting-flux dominated jet with a large emission radius,
a small length of the magnetic reconnection region,
or a low minimum energy of injected electrons would
prefer to form a Band-like radiation spectrum.

The radiation spectrum of the prompt emission is generally phenomenologically fitted
with a smoothly exponential-joint broken power-law, i.e., Band function.
However, it is found that there is another spectral break below the peak energy in some bright bursts (\citealp{Zheng_W-2012-Shen_RF-ApJ...751...90Z}; \citealp{Oganesyan_G-2017-Nava_L-ApJ...846..137O}; \citealp{Oganesyan_G-2018-Nava_L-A&A...616A.138O}).
Recently, \cite{Ravasio_ME-2019-Ghirlanda_G-A&A...625A..60R} systematically searched for this break in ten brightest long bursts from the $Fermi$/GBM catalog.
They found that there was convincing evidence of an additional spectral break at a range around $10\sim100$~keV in eight out of these ten bursts.
These features are usually seen as the signature of synchrotron radiation in the prompt emission (\citealp{Burgess_JM-2014-Preece_RD-ApJ...784...17B}; \citealp{Burgess_JM-2015-Ryde_F-MNRAS.451.1511B}; \citealp{Zhang_BB-2016-Uhm_ZL-ApJ...816...72Z}).
The standard fast-cooling radiation spectrum can be described with a double smoothly connected broken power-law (\citealp{Ravasio_ME-2018-Oganesyan_G-A&A...613A..16R}) with two spectral breaks associated with $\gamma'_{\rm c}$ and $\gamma'_{\rm m}$, respectively.
Because these two spectral breaks is fairly close to each other, the $\gamma'_{\rm c}$ should be $\sim 0.3\gamma'_{\rm m}$ for these bursts (\citealp{Oganesyan_GN-2019-Nava_L-A&A...628A..59O}).
It may be contrived in the standard synchrotron emission model.
We suggest that these spectrum features can be well explained in our scenario.
There are three main advantages of our scenario compared with the standard one.
(1) The lowest spectral break stems from the electrons with $\gamma'_{\rm e}=\gamma'_{\rm lb}$ satisfying ${\left[ {d{{n'}_{\rm{e}}}({{\gamma '}_{\rm{e}}})/d{{\gamma '}_{\rm{e}}}} \right]_{\gamma'_{\rm lb}}}=-1/3$ rather than $\gamma'_{\rm c}$.
This is because the synchrotron radiation of the electrons with $\gamma'_{\rm e}<\gamma'_{\rm lb}$ is
generally masked by the low-energy regime of the synchrotron emission from an individual electron
for a bump-shape electron spectrum.
It means that there is no need to take the $\gamma'_{\rm c}$ close to the $\gamma'_{\rm m}$ in our scenario, which may help to alleviate the aforementioned conflict in the standard synchrotron emission model.
(2) Our radiation spectrum is smoother compared with a double broken power-law
and would be more appropriate to describe the observed radiation spectrum.
(3) The first and second photon spectral indexes in the standard fast-cooling pattern are fixed at $-2/3$ and $-3/2$, respectively.
However, in our scenario, the second photon index can be different from $-3/2$.
It can vary between $-3/2$ and $-2/3$, which depends on
how fast the increase of $Q$ is or how strong the synchrotron cooling is.
We note that the above three advantages are all related to the bump-shape feature
in the low-energy regime of electron spectrum.
Therefore, we suggest that the Band radiation spectrum found in GRBs may be the synchrotron emission of the electrons with a bump-shape distribution in its low-energy regime.

\acknowledgments
We thank the anonymous referee of this work for beneficial suggestions that improved the paper.
We also thank Shen-Shi Du, Hao-Yu Yuan, and Zhao Zhang for the useful discussions and suggestions.
This work is supported by
the National Natural Science Foundation of China
(grant Nos. 11773007, 11533003, 11673006, 11851304, U1731239, U1938201),
the Guangxi Science Foundation (grant Nos. 2018GXNSFFA281010, 2017AD22006, 2018GXNSFGA281007, 2016GXNSFFA380006),
and the Innovation Team and Outstanding Scholar Program in Guangxi Colleges.

\clearpage
\begin{figure}
\begin{center}
\begin{tabular}{ccc}
\includegraphics[angle=0,scale=0.25,trim=60 0 60 0]{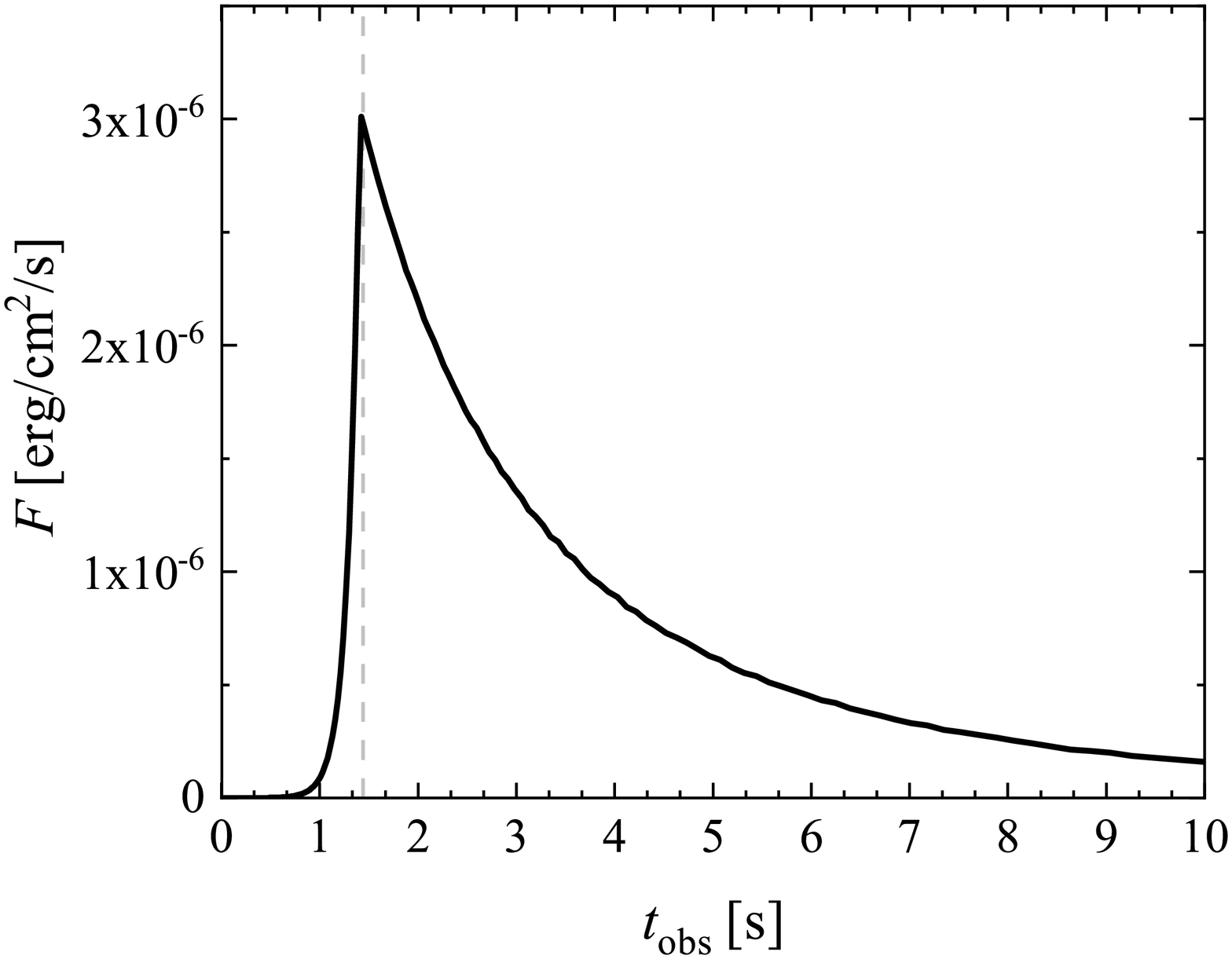} &
\includegraphics[angle=0,scale=0.263,trim=75 0 75 0]{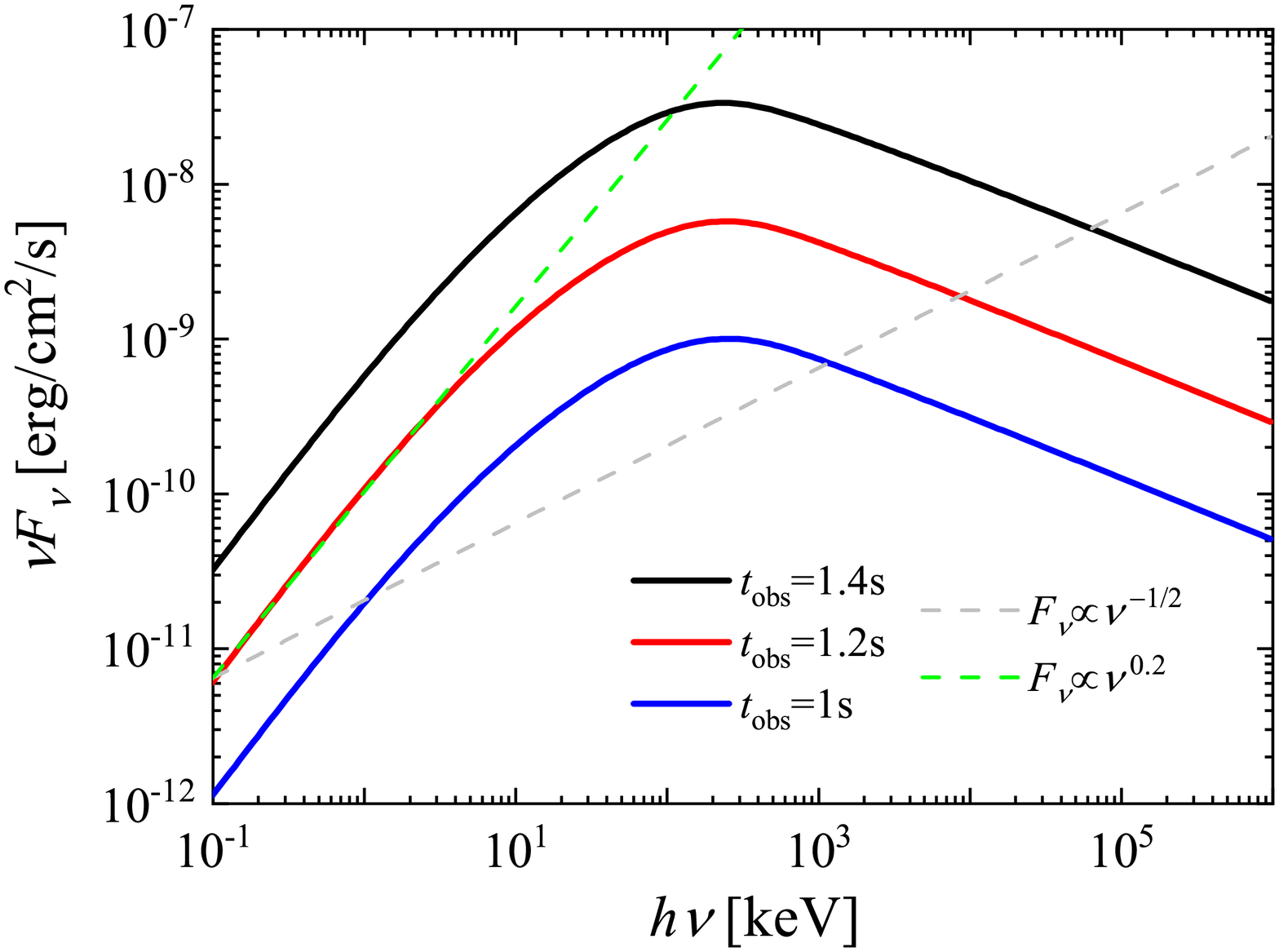} &
\includegraphics[angle=0,scale=0.25,trim=60 0 60 0]{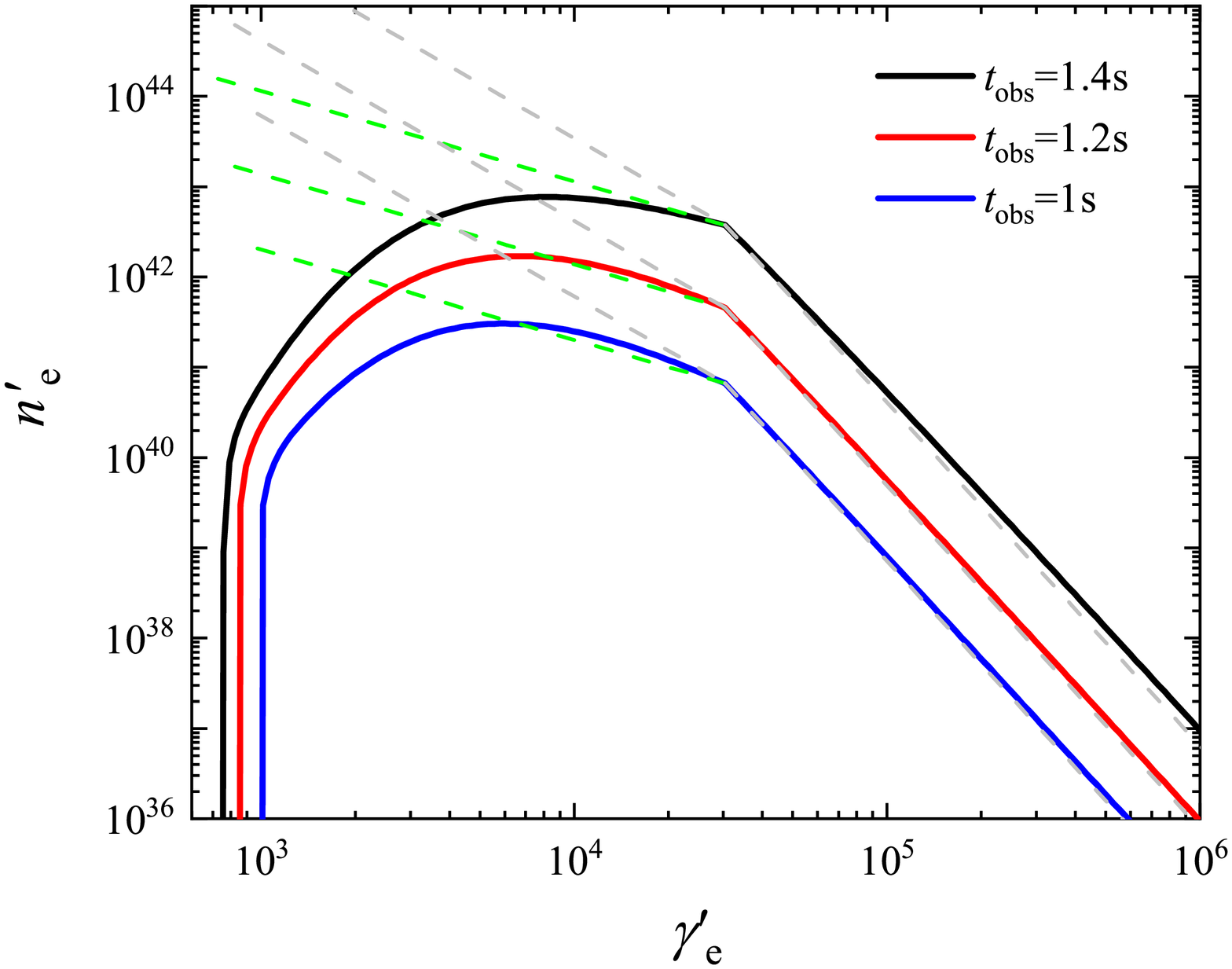} \\
\end{tabular}
\end{center}
\caption{The light curves (left panel), radiation spectra (middle panel), and electron spectra (right panel)
from the situation with an EXP-$Q$.
Here, the vertical gray dashed lines in the left panel indicate the peak time of the light curve.
In the middle panel, the blue, red, and black lines
plot the radiation spectra at the observer time $t_{\rm obs}=1$~s, $1.2$~s, $1.4$~s, respectively.
For comparison, we also plot the radiation spectra with $F_{\nu}\propto \nu^{-1/2}$ (gray dashed line)
and $F_{\nu}\propto \nu^{\alpha+1}$ with $\alpha=-0.8$ (green dashed line) in this panel.
In the right panel,
the electron spectra from the jet flow located at $\theta=0$ and observed at $t_{\rm obs}=1$~s, $1.2$~s, $1.4$~s
are plotted with the blue, red, and black lines, respectively.
In this panel, the gray dashed lines plot the electron spectra suffered from the standard fast synchrotron cooling and the green dashed lines plot the electron spectra required to produce $F_{\nu}\propto \nu^{0.2}$ radiation spectrum.
}\label{MyFigA}
\end{figure}
\begin{figure}
\begin{center}
\begin{tabular}{ccc}
\includegraphics[angle=0,scale=0.25,trim=60 0 60 0]{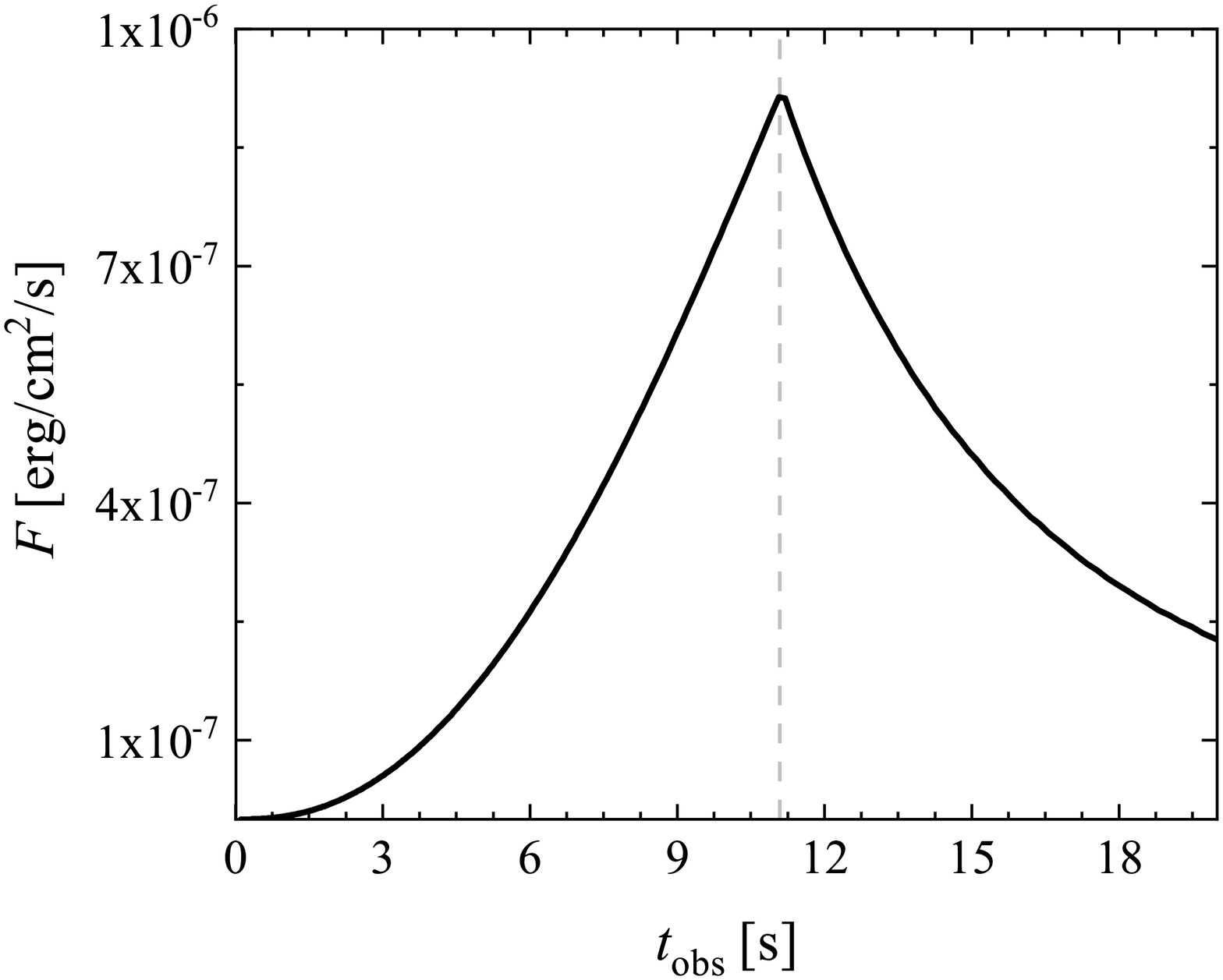} &
\includegraphics[angle=0,scale=0.263,trim=75 0 75 0]{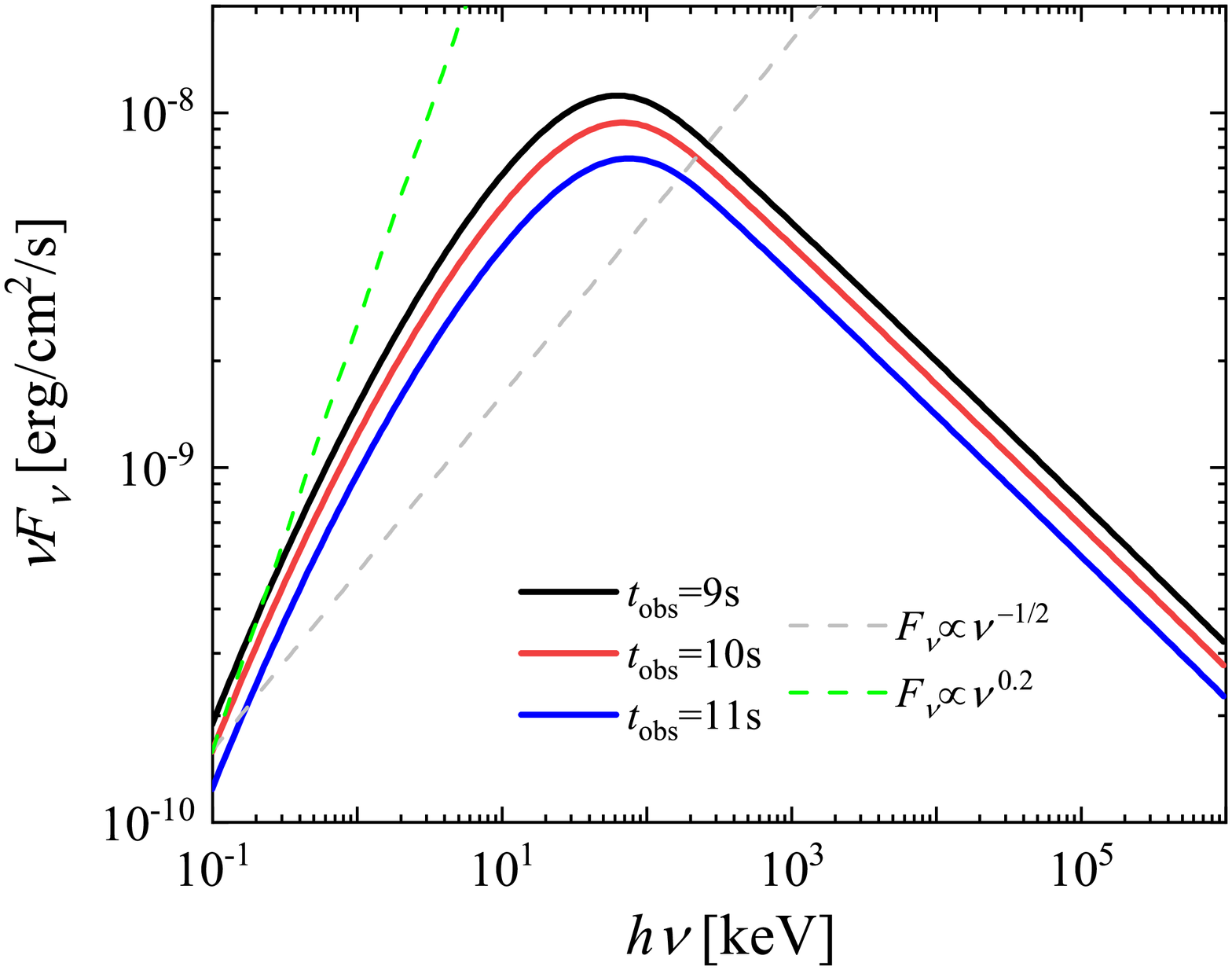} &
\includegraphics[angle=0,scale=0.25,trim=60 0 60 0]{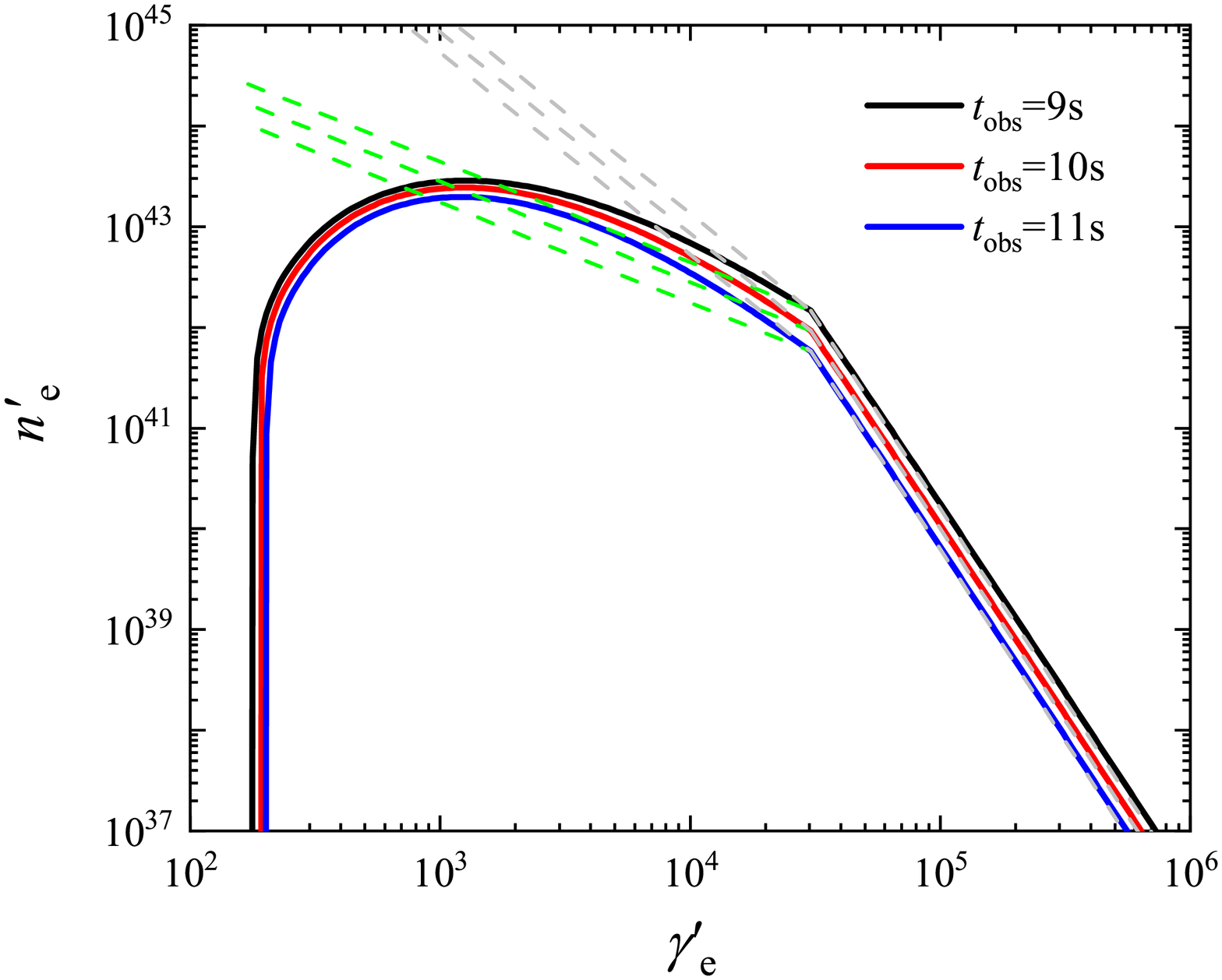} \\
\includegraphics[angle=0,scale=0.25,trim=60 0 60 0]{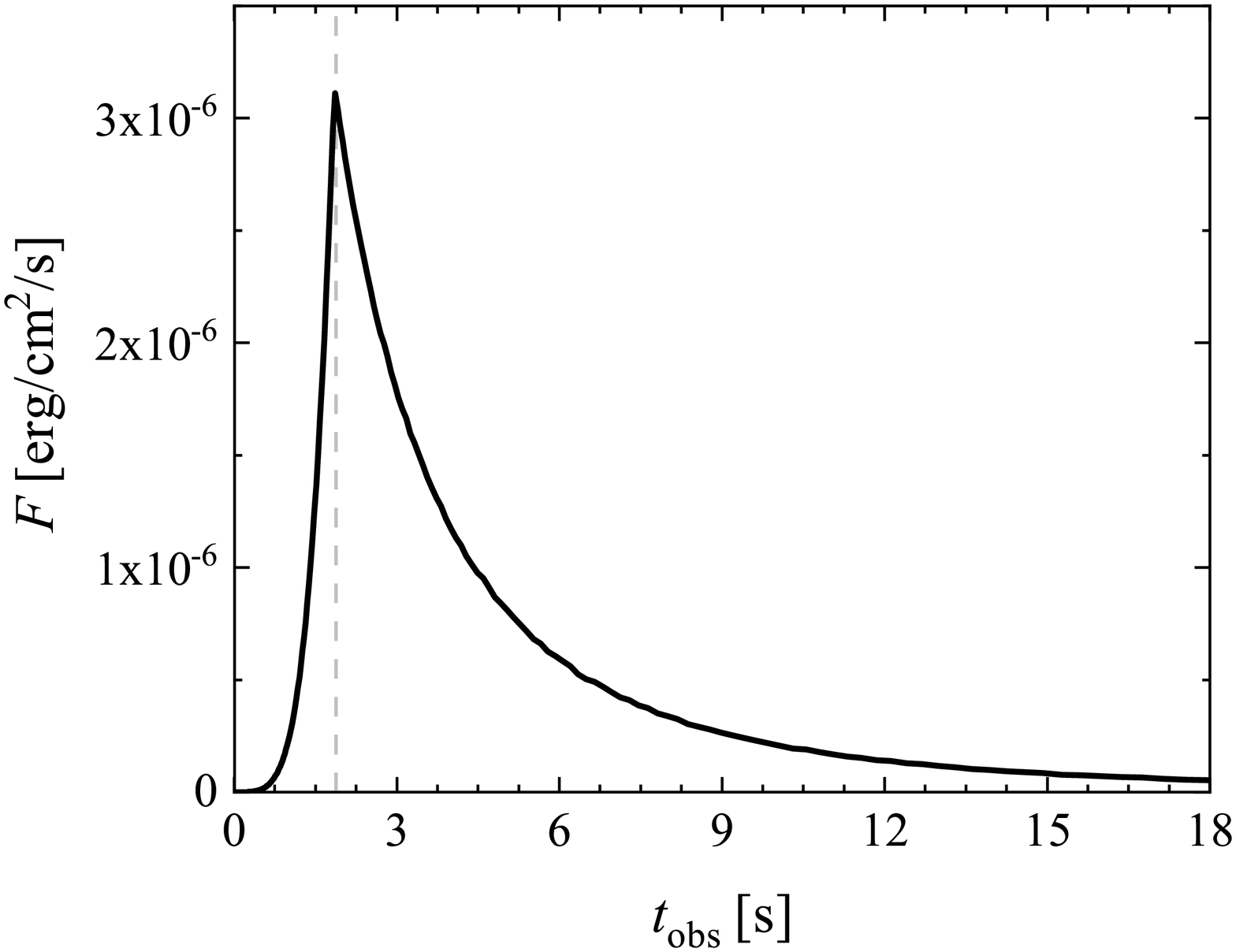} &
\includegraphics[angle=0,scale=0.263,trim=75 0 75 0]{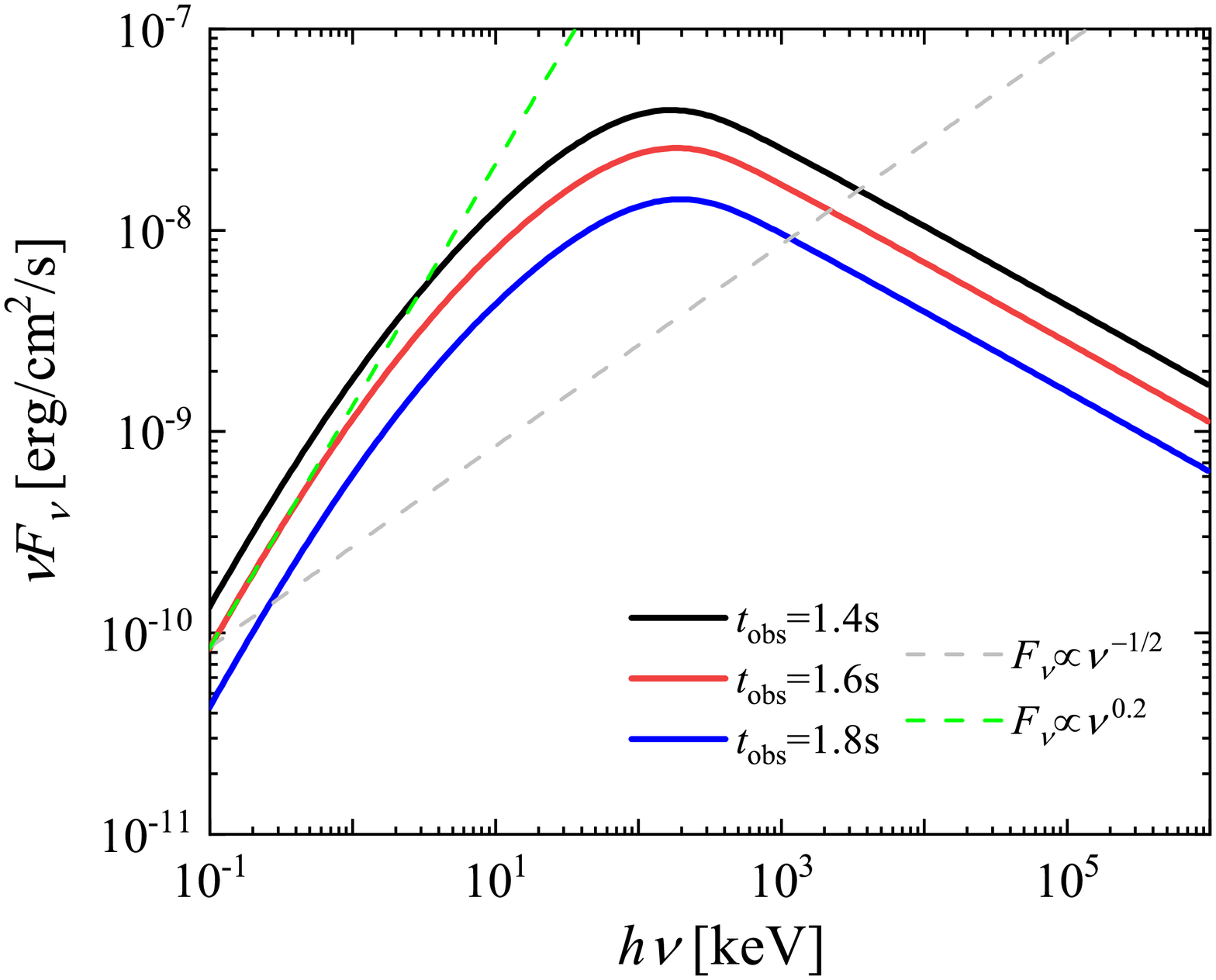} &
\includegraphics[angle=0,scale=0.25,trim=60 0 60 0]{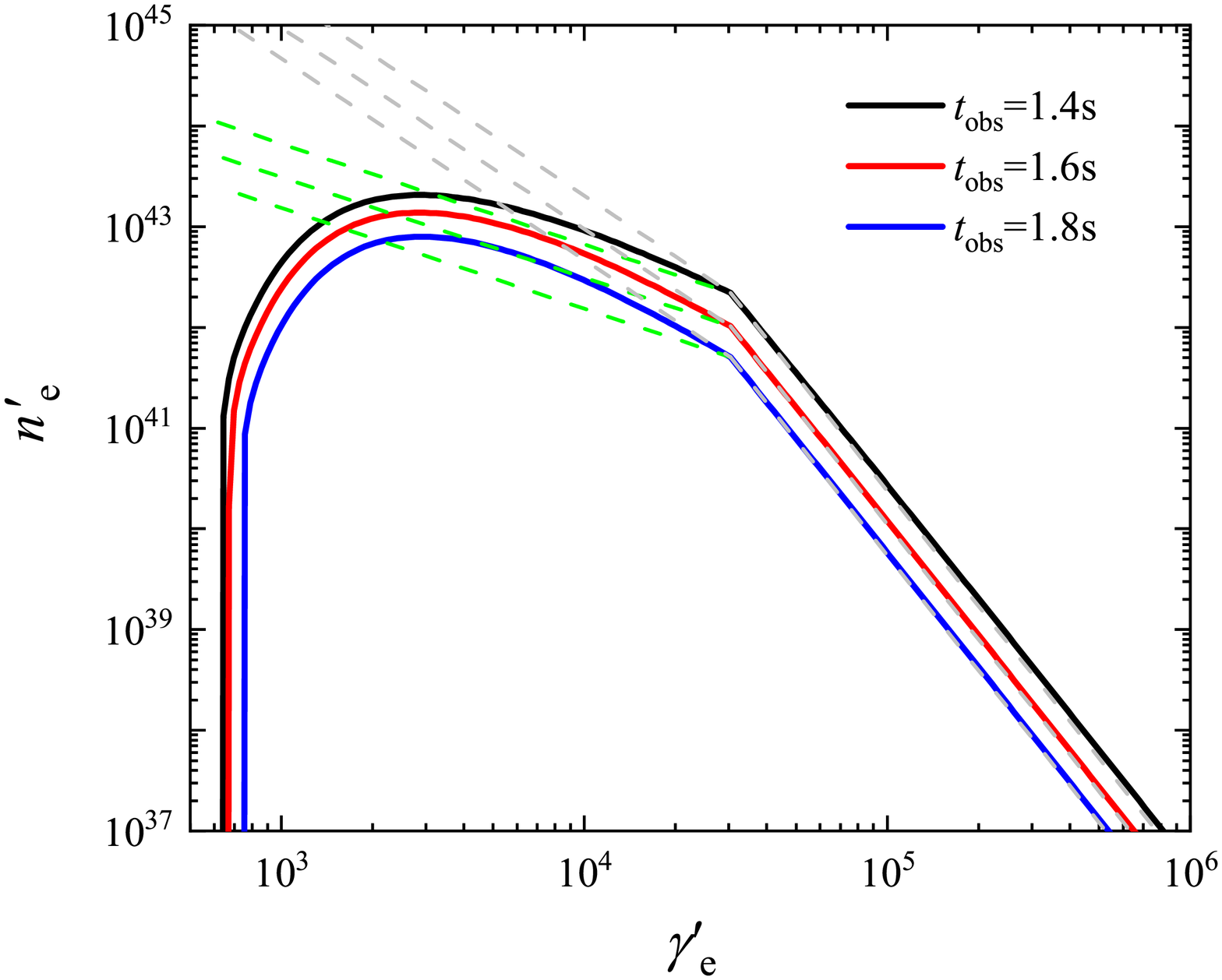} \\
\end{tabular}
\end{center}
\caption{The light curves (left panel), radiation spectra (middle panel), and electron spectra (right panel) from the situation with a PL-$Q$, where $\xi=2$and $\xi=4$ are adopted in the upper and bottom panels, respectively. The solid lines with blue, red, and black color
are the results corresponding to the observer time $t_{\rm obs}=1.4$~s, $1.6$~s, and $1.8$~s, respectively.
The meanings of other lines are the same as that of Figure~\ref{MyFigA}.}
\end{figure}
\begin{figure}
\begin{center}
\begin{tabular}{ccc}
\includegraphics[angle=0,scale=0.25,trim=75 0 75 0]{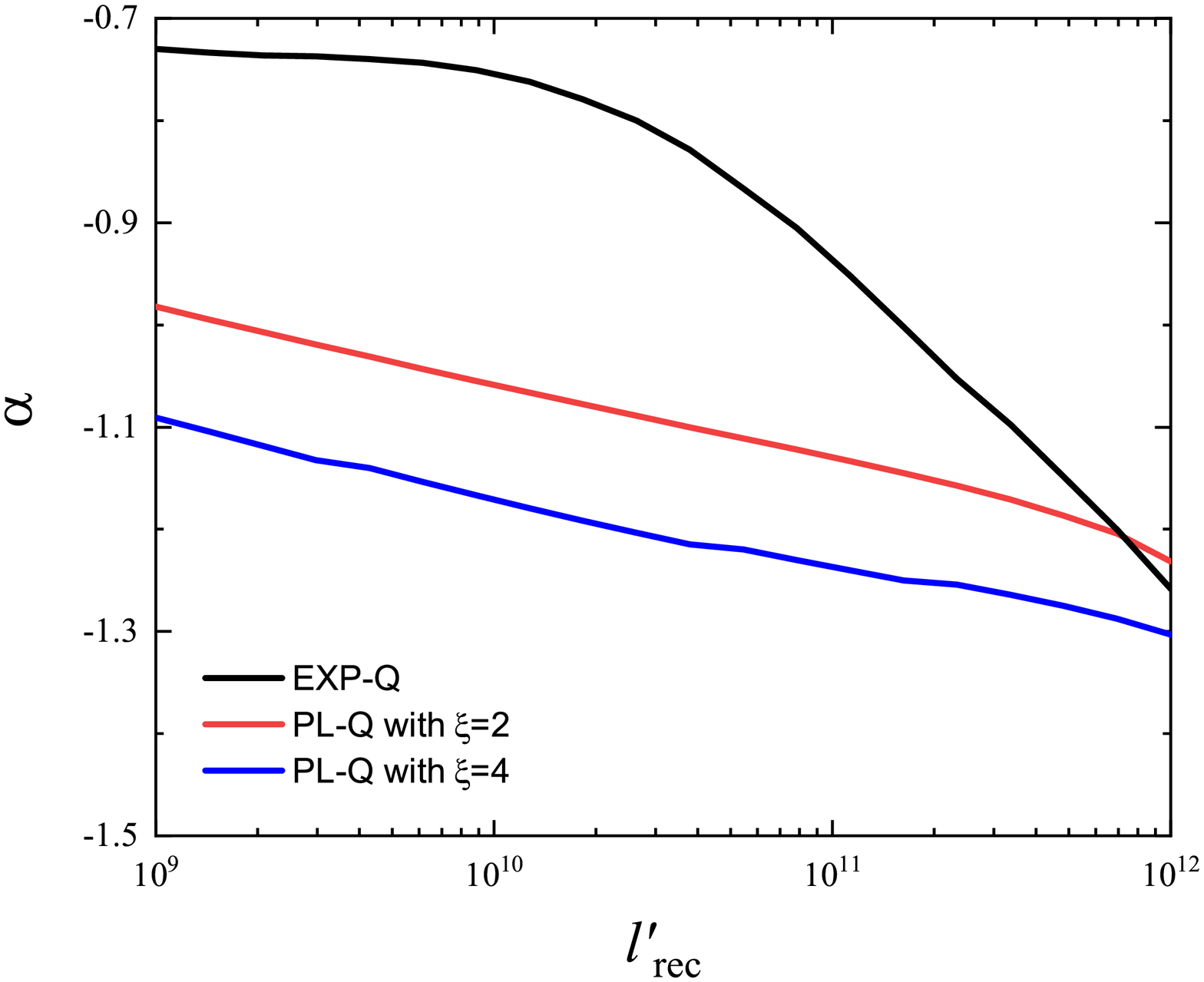} &
\includegraphics[angle=0,scale=0.25,trim=60 0 60 0]{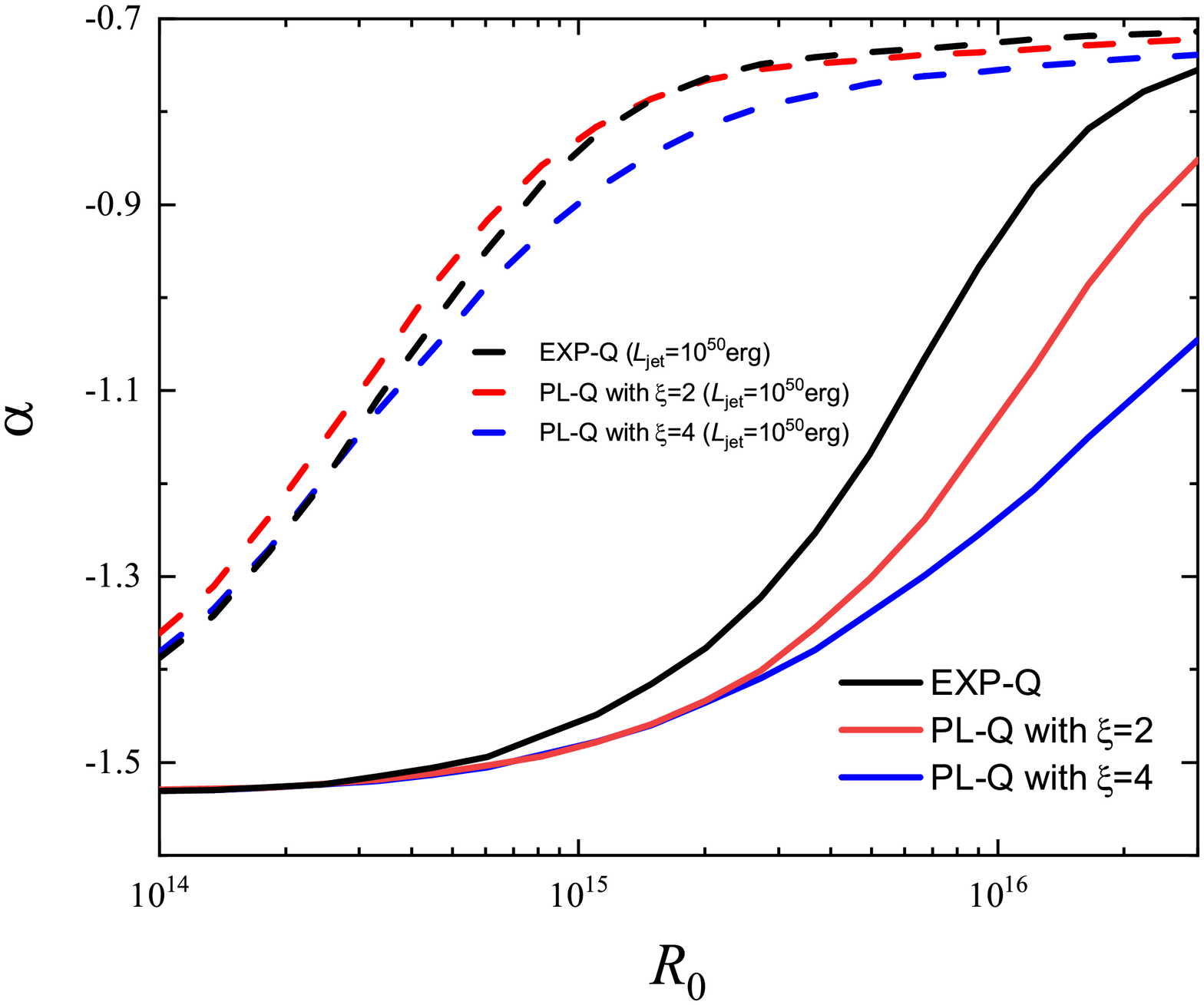} &
\includegraphics[angle=0,scale=0.25,trim=60 0 60 0]{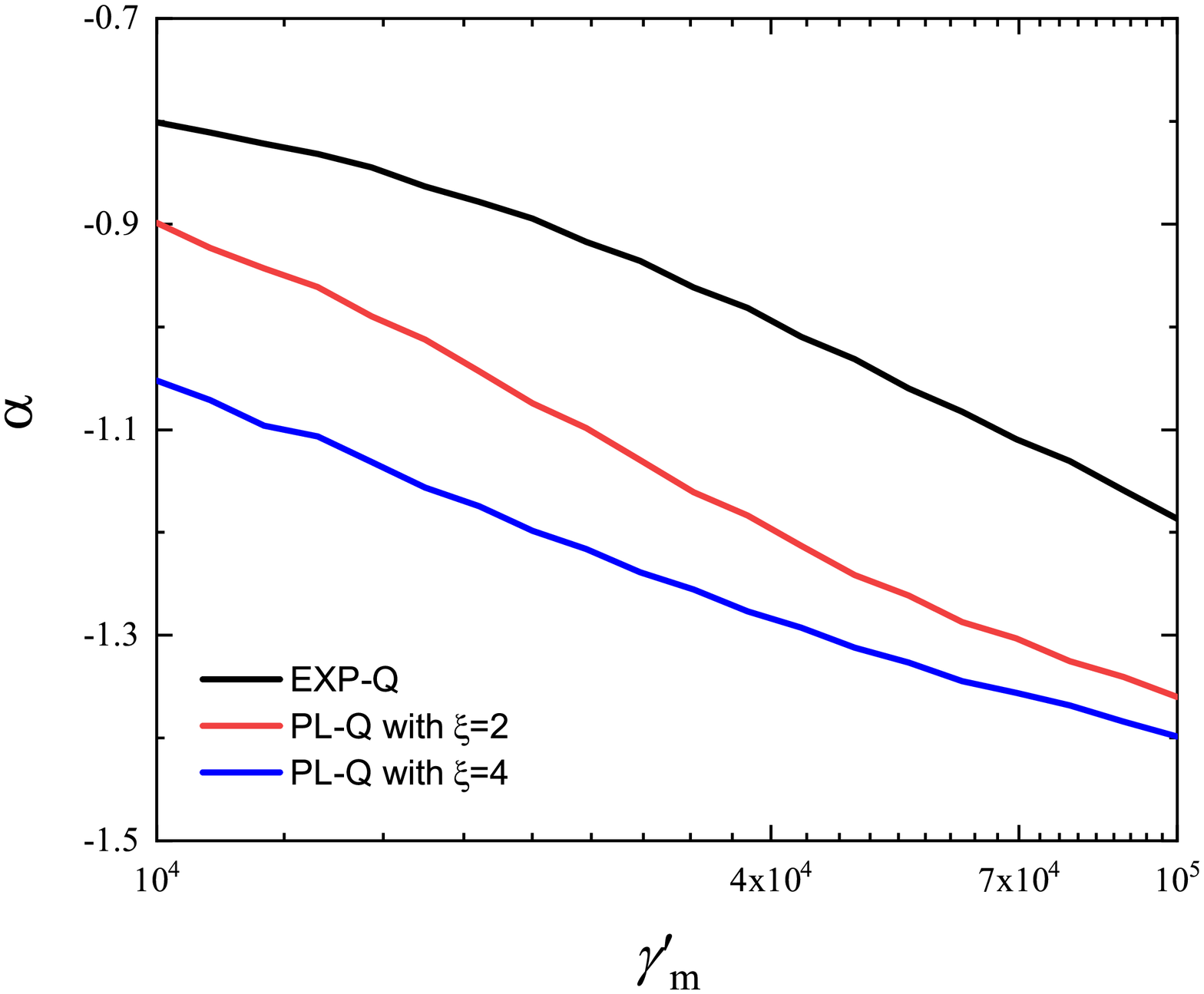}\\
\end{tabular}
\end{center}
\caption{The dependence of $\alpha$ on $l'_{\rm rec}$ (left panel), $R_0$ (middle panel), and $\gamma'_{\rm m}$ (right panel), respectively.
The situations by involving an EXP-$Q$, PL-$Q$ with $\xi=2$, and PL-$Q$ with $\xi=4$
are shown with black, red, and blue lines, respectively.
The dashed lines in the middle panel plot the dependence of $\alpha$ on $R_0$
in the situation with $L_{\rm jet}=10^{50}\, \rm erg\cdot s^{-1}$.
}
\end{figure}

\end{document}